\documentclass{PoS}
\usepackage{cite} 

\title{BESIII: ``charming'' physics at an e$^+$e$^-$ collider machine}

\ShortTitle{BESIII: ``charming'' physics}

\author{\speaker{Johan Messchendorp} for the BESIII Collaboration\\
        KVI/University of Groningen, Zernikelaan 25, 9747 AA Groningen, The Netherlands.\\
        E-mail: \email{messchendorp@kvi.nl}}

\abstract{Despite the successes of the standard model, the non-perturbative dynamics of the strong interaction are not fully understood yet. Charmonium spectroscopy serves as an ideal tool to shed light on the dynamics of the strong interaction such as quark confinement and the generation of hadron masses. The BESIII collaboration studies extensively the strong interaction and various aspects that could shed light on physics beyond the standard model via copious e$^+$e$^-$ collisions at the BESIII/BEPCII facility in Beijing, China, in the charmonium mass regime. We present a few of the recent results with the emphasis on charmonium spectroscopy studies using 106$\times$10$^6$ $\psi^\prime$ events.}

\FullConference{The 13th International Conference on B-Physics at Hadron Machines - Beauty2011,\\
		April 04-08, 2011\\
		Amsterdam, The Netherlands}

\begin{document}

\section{Introduction}

\begin{figure} \includegraphics[width=\textwidth]{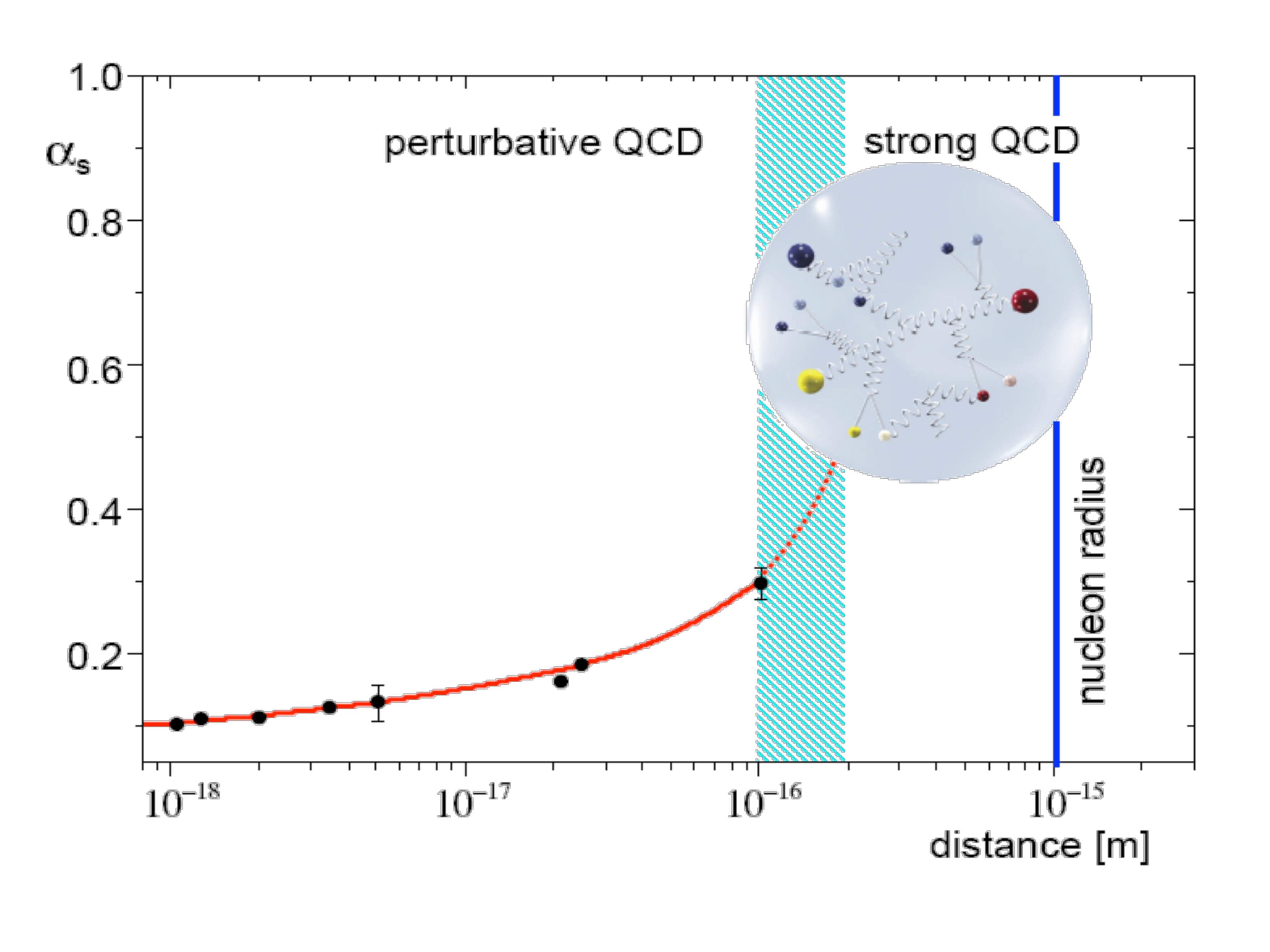}
\caption{The strong coupling constant, $\alpha_S$, as a function of the distance scale. Towards larger distances, QCD becomes non-perturbative, and gives rise to spectacular phenomena such as the generation of hadron masses.}
\label{fig_strong_coupling}
\end{figure}

The fundamental building blocks of Quantum Chromodynamics (QCD) 
are the quarks which interact with each other by exchanging gluons. QCD is well 
understood at short-distance scales, much shorter than the 
size of a nucleon ($<$~10$^{-15}$~m). In this regime, the basic 
quark-gluon interaction is sufficiently weak. In fact, many processes 
at high energies can quantitatively be described by perturbative QCD.
Perturbation theory fails when the distance among quarks becomes 
comparable to the size of the nucleon. Under these conditions, in the 
regime of non-perturbative strong QCD, the force among the quarks 
becomes so strong that they cannot be further separated (see illustration in Fig.~\ref{fig_strong_coupling}). As a 
consequence of the strong coupling, we observe the relatively heavy 
mass of hadrons, such as protons and neutrons, which is two orders of 
magnitude larger than the sum of the masses of the individual quarks. 
This quantitatively yet-unexplained behavior is related to the 
self-interaction of gluons leading to the formation of gluonic 
flux tubes connecting the quarks. As a consequence, quarks have 
never been observed as free particles and are confined within hadrons, 
i.e. the baryons containing three valence quarks or mesons containing 
a quark-antiquark pair. 

The level scheme of lower-lying bound $\bar c$$c$ states, charmonium, 
is very similar to that of positronium. These charmonium states
can be described fairly well in terms of heavy-quark potential models.
Precision measurements of the mass and width of the charmonium spectrum
give, therefore, access to the confinement potential in QCD.  
In addition, the various charmonium states with well-defined spin and parity
serve as ideal systems to study via their decay modes the validity of perturbative 
QCD and to probe the light-quark sector as well. 

\section{The BESIII experiment}

BEPCII is a two-ring e$^+$e$^-$ collider designed for a peak luminosity of 10$^{33}$~cm$^{-2}$s$^{-1}$ at a beam current of 0.93~A. The cylindrical core of the BESIII detector~\cite{bes_bepc} consists of a helium-gas-based drift chamber, a plastic scintillator time-of-flight system, and a CsI(Tl) 
electromagnetic calorimeter, all enclosed in a superconducting solenoidal magnet providing 
a 1.0~T magnetic field. The solenoid is supported by an octagonal flux-return yoke with resistive plate counter muon identifier modules interleaved with steel. The charged particle and photon acceptance is 93\% of 4$\pi$, and the charged particle momentum and photon energy resolutions at 1~GeV are 0.5\% and 2.5\%, respectively. Both the BEPCII facility and the BESIII detector are major upgrades of the BESII detector and the BEPC accelerator. The first collisions with the complete setup took place in July of 2008. The first physics production runs started in the first half of 2009. Already during writing of this paper, the amount of data samples collected for the $J/\psi$, $\psi^\prime$, and $\psi(3770)$ is significantly larger than that obtained by the CLEO collaboration, thereby reaching a new world record in statistics.

\section{Some recent results}

The BESIII collaboration has published so-far a variety of papers with many new results in
the field of charmonium spectroscopy and charmonium decays~\cite{hc_paper,gP_paper,PP_paper,ppbar_paper,X1835_paper,matrix_paper,4pi_paper,mixing_paper,gV_paper}. A number of new hadronic states were discovered or confirmed, and various decay properties were measured for the first time or with a better precision than published before. In addition, many data analyses are in an advanced stage and will lead to a rich set of new publications in the near future. Here, we show only a small fraction of the published results, thereby, illustrating the potential of the BESIII experiment.

\subsection{The $h_c$ charmonium state}

\begin{figure} \includegraphics[width=\textwidth]{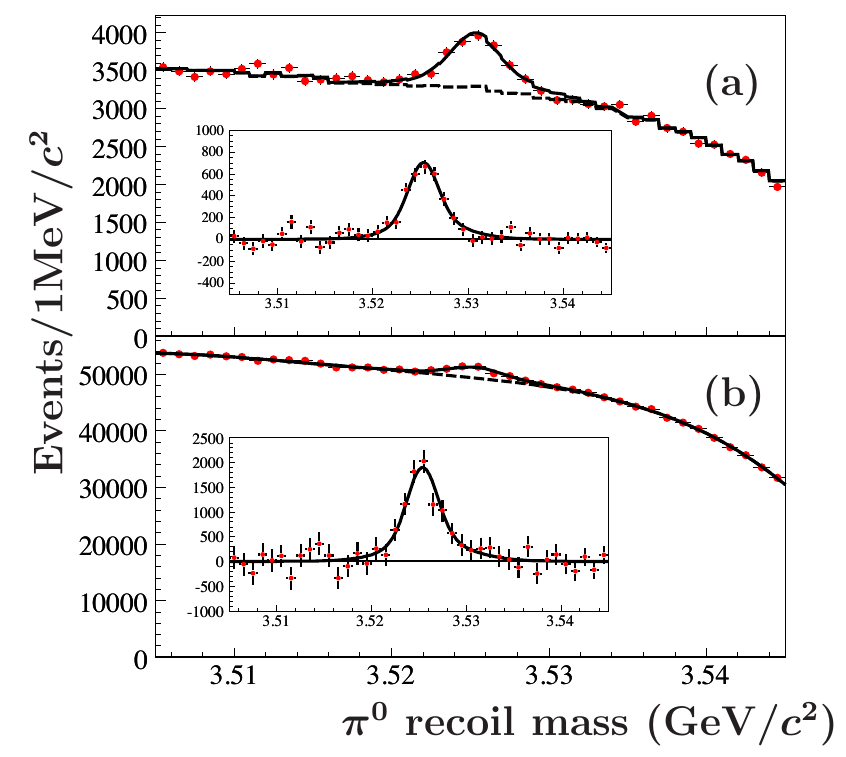}
\caption{(a) The $\pi^\circ$ recoil mass spectrum and fit for the E1-tagged analysis of $\psi^\prime \rightarrow \pi^\circ h_c$, $h_c\rightarrow \gamma \eta_c$; (b) the $\pi^\circ$ recoil mass spectrum and fit for the inclusive analysis of $\psi\ \rightarrow \pi^\circ h_c$. Fits are shown as solid lines, background as dashed lines. The insets show the background-subtracted spectra. Figures are taken from Ref.~\cite{hc_paper} and described there in more detail.}
\label{fig_hc}
\end{figure}

One of the important aspects related to quark confinement is the spin structure of the $q\bar q$ potential. The role of the spin-dependence in the hyperfine splitting of the $P$-waves is of particular interest. For this purpose, a precise measurement of the mass and decay channels of the singlet-$P$ resonance, $h_c$, is of extreme importance.
This state has been studied by the BESIII collaboration via the isospin-forbidden transition, 
$\psi^\prime\rightarrow \pi^\circ h_c$. The results of this analysis are shown in Fig.~\ref{fig_hc}. Clear signals have been observed for this decay with and without the subsequent radiative decay, $h_c\rightarrow \gamma\eta_c$. This has led to a measurement of the mass and the total width of the $h_c$ of $M=3525.40\pm0.13\pm0.18$~MeV/$c^2$ and $\Gamma=0.73\pm0.45\pm0.28$~MeV ($<$1.44~MeV at 90\% C.L.), respectively. Furthermore, for the first time the branching fractions of
the decays $\psi^\prime\rightarrow \pi^\circ h_c$ and $h_c\rightarrow\gamma\eta_c$ were determined and found to be $B(\psi^\prime\rightarrow \pi^\circ h_c)=(8.4\pm1.3\pm1.0)\times 10^{-4}$ and $B(h_c\rightarrow \gamma\eta_c)=(54.3\pm6.7\pm5.2)\%$, respectively. The measured $1P$ hyperfine mass splitting $\Delta M_{hf} \equiv <M(1^3P)>-M(1^1P_1) = -0.10\pm0.13\pm0.18$~MeV/$c^2$ is consistent with there being no strong spin-spin interaction. For a more detailed discussion, we refer to Ref.~\cite{hc_paper}.

\subsection{The radiative decays $\psi^\prime\rightarrow \gamma P$ with $P=\{\pi^\circ,\eta,\eta^\prime\}$}

The radiative decay of the $\psi^\prime$ to a pseudo-scalar meson, such as the $\pi^\circ$, $\eta$, and $\eta^\prime$, is of interest since it provides a system to study the two-gluon coupling to $c\bar c$ states, to study the $\eta - \eta^\prime$ mixing angle, and to probe the $\pi^\circ$ form factor in the time-like region.
Recently, the CLEO collaboration reported measurements for the decays of $J/\psi$, $\psi^\prime$, and $\psi^{\prime\prime}$ to $\gamma P$~\cite{cleo_gP_paper}, and no evidence for $\psi^\prime\rightarrow \gamma\eta$ or $\gamma\pi^\circ$ was found. With BESIII, the processes $\psi^\prime\rightarrow \gamma\pi^\circ$ and $\psi^\prime\rightarrow \gamma\eta$ are observed for the first time with signal significances of 4.6$\sigma$ and 4.3$\sigma$, respectively, and with branching fractions of $B(\psi^\prime\rightarrow\gamma\pi^\circ)=(1.58\pm0.40\pm0.13)\times 10^{-6}$
and $B(\psi^\prime\rightarrow \gamma\eta)=(1.38\pm0.48\pm0.09)\times 10^{-6}$. 
With a measured branching fraction, $B(\psi^\prime\rightarrow\gamma\eta^\prime)=(126\pm 3\pm 8)\times 10^{-6}$, the BESIII collaboration determined for the first time the ratio of the $\eta$ and $\eta^\prime$ production rates from $\psi^\prime$ decays, $R_{\psi^\prime}\equiv B(\psi^\prime\rightarrow \gamma\eta)/B(\psi^\prime\rightarrow \gamma\eta^\prime)=(1.10\pm0.38\pm0.07)$\%. This ratio is below the 90\% C.L. upper bound determined by the CLEO collaboration and, in contradiction to leading-order perturbative-QCD predictions, one order of magnitude smaller than the corresponding ratio for the $J/\psi$ decays, $R_{J/\psi}=(21.1\pm0.9)$\%. More details on the data analysis can be found in Ref.~\cite{gP_paper}.

\begin{figure} \includegraphics[width=\textwidth]{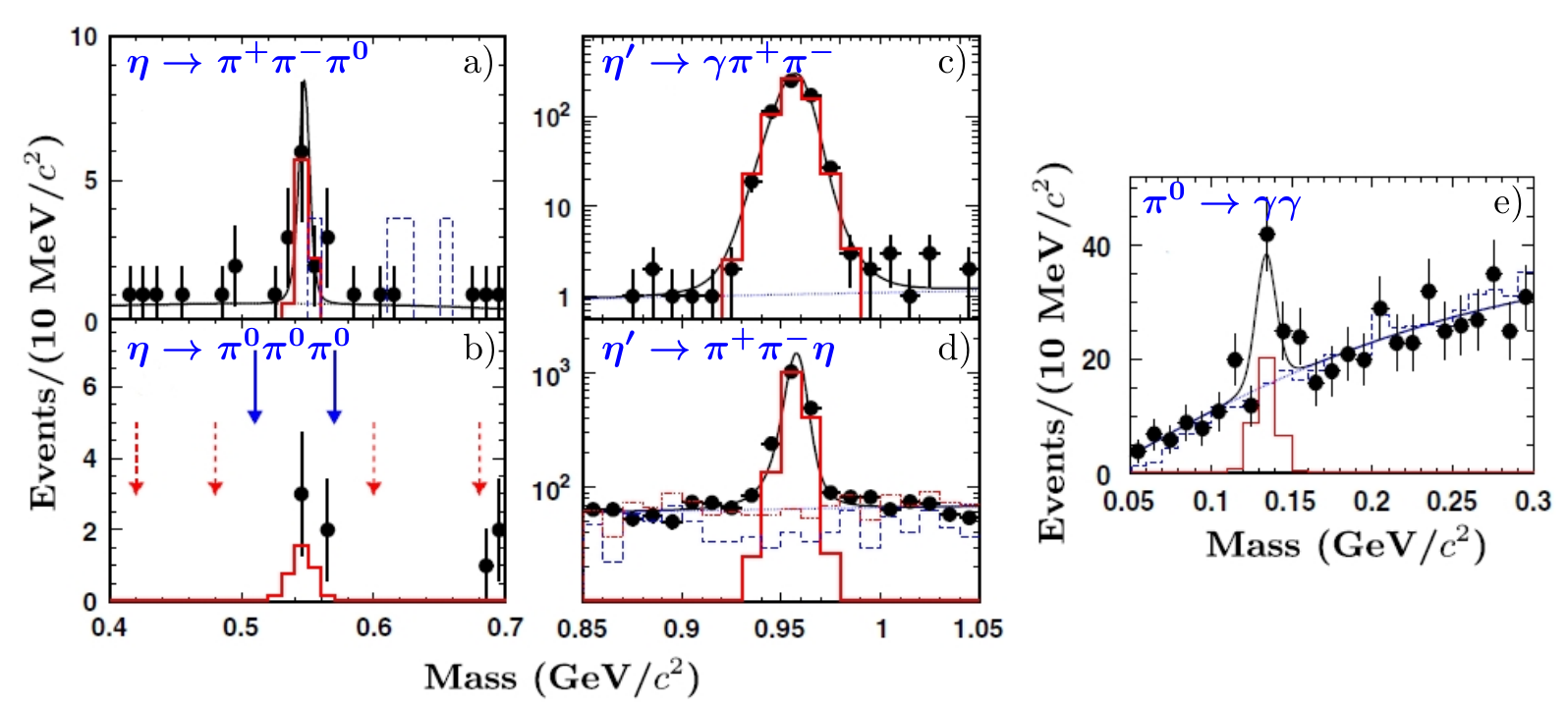}
\caption{Mass distributions of the pseudo-scalar meson candidates for $\psi^\prime\rightarrow \gamma P$:
a) $P=\eta(\rightarrow\pi^+\pi^-\pi^\circ)$; b) $P=\eta(\rightarrow 3\pi^\circ)$; c) $P=\eta^\prime(\rightarrow \gamma\pi^+\pi^-)$; d) $P=\eta^\prime(\rightarrow \pi^+\pi^-\eta(\rightarrow \gamma\gamma))$; d) $P=\pi^\circ(\rightarrow \gamma\gamma)$. For more details, we refer to Ref.~\cite{gP_paper}.}
\label{fig_gP}
\end{figure}

\subsection{The radiative decays $\chi_{cJ}\rightarrow \gamma V$ with $V=\{\rho,\omega,\phi\}$}

The copious number of $\psi^\prime$ events obtained with BESIII, allows one to study in great detail the decays of the $P$-wave $\chi_{cJ}$ states with $J=0, 1, 2$ via the E1 transition $\psi^\prime\rightarrow \gamma \chi_{cJ}$. One of the many physics channels that can be probed is the radiative decay of the $\chi_{cJ}$ states into light vector mesons, $\rho,\omega,\phi$. Similar as for the radiative decay of the $\psi^\prime$ into light pseudo-scalar mesons, the $\chi_{cJ}$ decays into light vector mesons provide access to the two-gluon coupling to $c\bar c$ states as well but with the difference, that the photon couples to one of the light quarks. Perturbative QCD calculations predict the radiative branching ratios into light vector mesons to be in the range of $(0.1-15)\times 10^{-6}$~\cite{pQCD_paper} depending on the total spin, $J$, of the $\chi_{cJ}$ state and the type of vector meson. With the available BESIII data, a variety of such branching ratios can be measured. For example, a recent analysis resulted for the first time in a branching ratio of the decay $\chi_{c1}\rightarrow\gamma \phi$ of $B=(25.8\pm 5.2\pm 2.3)\times 10^{-6}$~\cite{gV_paper}. Surprisingly, the measured branching ratio is about an order of magnitude larger than the predictions by perturbative QCD ($3.6\times 10^{-6}$), which could point to non-perturbative QCD effects playing an important role in these decays. Similar observations have been made for the decays $\chi_{c1}\rightarrow \gamma \rho^\circ$ and $\chi_{c1}\rightarrow \gamma \omega$. A recent analysis gave branching ratios of $B(\chi_{c1}\rightarrow \gamma\rho^\circ)=(228\pm 13\pm 22)\times 10^{-6}$ and $B(\chi_{c1}\rightarrow \gamma\omega)=(69.7\pm 7.2\pm 6.6)\times 10^{-6}$~\cite{gV_paper} which are both significantly larger than predictions by perturbative QCD, $B(\chi_{c1}\rightarrow \gamma\rho^\circ)=14\times 10^{-6}$ and $B(\chi_{c1}\rightarrow \gamma\omega)=1.6 \times 10^{-6}$~\cite{pQCD_paper}, and consistent with published data from CLEO~\cite{cleo_paper}. For a more detailed description of the BESIII analysis, we refer to Ref.~\cite{gV_paper}.

\section{Summary and outlook}

The BESIII experiment at the BEPCII facility addresses a wide range of
topics in the field of QCD and searches for new physics beyond 
the Standard Model via e$^+$e$^-$ collisions. BESIII is fully
operational and is producing a rich set of data for physics studies.
At present the BESIII collaboration has collected a record on
statistics on $J/\psi$, $\psi^\prime$, and $\psi(3370)$ charmonium
states. These data are being exploited to provide precision 
measurements with a high discovery potential in charmonium spectroscopy, 
charmonium decays into light hadrons, and open charm production. 
Only a selection of data could be presented in this paper.

Very recently, data have been taken at a center-of-mass energy of 
4010~GeV. These data could give new insights in our understanding of the recently discovered
XYZ states and will allow the collaboration to explore the field of $D_s$ physics.
For the near future, a $\tau$ -mass scan has been scheduled. Furthermore, additional data will be taken at the $J/\psi$ and $\psi^\prime$ mass.

\end{document}